\documentclass[apj,graphicx]{emulateapj}
\usepackage{apjfonts}

\newcommand\cen{Cen~X-4}
\newcommand\suz{{\it Suzaku}}
\newcommand\xmm{{\it XMM-Newton}}
\newcommand\cxo{{\it Chandra}}
\newcommand\aql{Aql~X-1}
\newcommand\asca{{\it ASCA}}

\begin{document}

\title{Quiescent X-ray emission from Cen~X-4: a variable thermal component}
\shortauthors{Cackett et al.}
\shorttitle{Thermal variability in \cen}

\author{Edward~M.~Cackett\altaffilmark{1,4}, Edward~F.~Brown\altaffilmark{2}, Jon~M.~Miller\altaffilmark{1}, Rudy~Wijnands\altaffilmark{3}}

\email{ecackett@umich.edu}

\affil{\altaffilmark{1} Department of Astronomy, University of Michigan, 500 Church St, Ann Arbor, MI 48109-1042, USA}
\affil{\altaffilmark{2} Department of Physics \& Astronomy, National Superconducting Cyclotron Laboratory, and the Joint Institute for Nuclear Astrophysics, Michigan State University, East Lansing, MI 48824, USA}
\affil{\altaffilmark{3} Astronomical Institute `Anton Pannekoek', University of Amsterdam, Science Park 904, 1098 XH Amsterdam, the Netherlands}

\altaffiltext{4}{Chandra Fellow}

\begin{abstract}
The nearby neutron star low-mass X-ray binary, \cen, has been in a quiescent state since its last outburst in 1979.  Typically, quiescent emission from these objects consists of thermal emission (presumably from the neutron star surface) with an additional hard power-law tail of unknown nature.  Variability has been observed during quiescence in \cen{} on both timescales as short as hundreds of seconds and as long as years.  However, the nature of this variability is still unknown.  Early observations seemed to show it was all due to a variable hard X-ray tail. Here, we present new and archival observations that contradict this.  The most recent {\it Suzaku} observation of \cen{} finds it in a historically low state, a factor of 4.4 fainter than the brightest quiescent observation.  As the spectrum during the brightest observation was comprised of approximately 60\% from the thermal component and 40\% from the power-law component, such a large change cannot be explained by just power-law variability.  Spectral fits with a variable thermal component fit the data well, while spectral fits allowing both the column density and the power-law to vary do not, leading to the conclusion that the thermal component must be variable.  Interestingly, we also find that the thermal fraction remains consistent between all epochs, implying that the thermal and power-law fluxes vary by approximately the same amount.  If the emitting area remains unchanged between observations, then the effective surface temperature must change.  Alternatively, if the temperature remains constant, then the emitting area must change.  The nature of this thermal variability is unclear, but may be explained by variable low-level accretion.
\end{abstract}

\section{Introduction}
Neutron star low-mass X-ray binaries (LMXBs) are often transient.  During outbursts they accrete at a significant fraction of the Eddington luminosity, yet these outbursts tend to only last a few months in most cases.  Thus, for the majority of the time, these sources are in a quiescent state where the X-ray luminosity is significantly fainter.  The X-ray spectra of neutron star LMXBs during quiescence usually consists of two components -- a soft thermal component and a harder power-law component \citep[see e.g.][for a review]{campanaetal98a}.  The soft component is usually interpreted as thermal emission from the neutron star surface. The neutron star is heated during outbursts as compression by the accreted material causes nuclear reactions to occur deep in the crust. This heat is then radiated thermally during quiescence \citep{BBR98}.  The hard power-law component is less well understood and may be associated with residual accretion, or pulsar shock emission \citep[e.g.,][]{campanaetal98b,campanastella00,menoumcclintock01}.

Accurately measuring neutron star radii is vital for discriminating between the large range of possibilities for the dense matter equation of state \citep[e.g.,][]{lattimer07}.  There are a number of potential methods for constraining neutron star radii, for instance using X-ray bursts \citep[e.g.,][]{ozel09,guver10}, quasi-periodic oscillations \citep{miller98} or relativistic Fe K emission lines \citep{cackett08b}.  However, one of the most promising methods for measuring a neutron star radius, $R$, uses the thermal emission from quiescent neutron stars because as with any blackbody-like emission $f \propto R^2/D^2$ (where $f$ is the source flux, and $D$ the distance).   While many quiescent neutron stars are too faint for this method to produce constraining radius measurements, modest constraints have been possible for several objects \citep[e.g.,][]{heinke_eos_03,heinke06,webb07}.  Future X-ray telescopes, such as the {\it International X-ray Observatory}, will allow accurate radius measurements from many quiescent neutron stars in the Galaxy.

Nevertheless, there may be a potential problem.  Quiescent emission from neutron stars has been seen to be variable\citep{campana97,rutledgeetal01a,rutledge02,campanastella03,campanaetal04,cackett05}.  In most cases this variability comes from observing sources at different epochs, and can be explained away by variations in the power-law, for instance due to changes in residual accretion rate \citep[e.g.,][]{cackett05}.  However, in the case of \cen{} \citep{campanaetal04}, variability was observed {\it during} an \xmm{} observation, and the nature of the variability remains unclear.  If it is due to changes in temperature of the thermal component, this poses problems for neutron star radius measurements.  A mechanism for any short-timescale temperature change is uncertain, especially under the standard deep crustal heating picture.  Thus, this variability has important implications for radius measurements using thermal emission from quiescent neutron stars. 

\subsection{Variability in quiescence}
Several explanations have been used to describe the X-ray variability of quiescent neutron star LMXBs.  In \aql{} \citet{rutledge02} found that the temperature of the thermal component varied between observations.  Conversely, an analysis of the same data by \citet{campanastella03} came to an opposing conclusion, suggesting that the variability could be attributed to correlated changes in the column density and the slope of the power-law.  Moreover, an analysis of two observations of the neutron star LMXB in the globular cluster NGC~6440 led \citet{cackett05} to suggest that it was the power-law component varying in that object.

Probably the best studied quiescent neutron star is \cen{}, and multiple X-ray missions have observed it in quiescence.  \cen{} is one of the nearest known quiescent neutron star LMXB \citep[$D=1.2\pm0.3$ kpc;][]{chevalier89}.  X-ray variability of \cen{} in quiescence has been known about for 10 years \citep{campana97}.  It has been seen to be variable over timescales of 5 years, varying by 40\% \citep{rutledgeetal01a}, over a period of 3 days, varying by a factor of $\sim$3 \citep{campana97} and most recently on timescales as short as 100 seconds with an rms variability of 45\% \citep{campanaetal04}.  The spectrum of \cen{} clearly has both a thermal component and a power-law (with an index in the range 1-2) present, yet, studies so far have been inconclusive as to whether it is the temperature of the thermal component, the spectral index of the power-law or the strength of the power-law component that is variable. 

\citet{rutledgeetal01a} find a 40\% change in the quiescent luminosity of \cen{}  between {\it ASCA} and {\it Chandra} observations $\sim$5 years apart, which they attribute to the power-law component.  However, on shorter timescales it is unclear whether this interpretation holds.  \citet{campanaetal04} performed a detailed analysis of the quiescent spectrum of \cen, finding variability on 100 s timescales throughout a $\sim$50 ks \xmm{} observation.  A color-color analysis was not conclusive as to the source of the variability, thus, they extracted three separate spectra from the observation depending on the count rate. Based on an analysis of the low, medium and bright count rate spectra they remained unable to conclusively determine which component led to the variability.

In this paper, we present an analysis of new {\it Suzaku} data of \cen, as well as, analyzing two archival {\it Chandra} and \xmm{} observations of this source.  From spectral fitting, we show that the thermal component in \cen{} must be variable.

\section{Data Reduction}

In Table~\ref{tab:obs} we detail all the observations analyzed here.  Below we discuss the data reduction for each specific telescope.

\begin{deluxetable*}{lccccc}
\tablecolumns{6}
\tablewidth{0pc}
\tablecaption{Observations of Cen~X-4}
\tablehead{Mission & Obs. ID & Short name & Start date & Exposure time(ks) & Reference}
\startdata
\asca & 41008100   & ASCA & 27/02/1994 & 39  & 1, 2, 3, 4  \\
\cxo  & 713        & CXO1 & 23/06/2000 & 10  & 4 \\
\xmm  & 0067750101 & XMM1 & 20/08/2001 & 53  & 5 \\
\xmm  & 0144900101 & XMM2 & 01/03/2003 & 78  & 6 \\
\cxo  & 4576       & CXO2 & 21/06/2004 & 10  & 6 \\
\suz  & 403057010  & SUZ  & 16/01/2009 & 147 & 6 
\enddata
%\tablecomments{}
\tablerefs{
(1) \citealt{asai96},
(2) \citealt{asai98},
(3) \citealt{rutledgeetal99},
(4) \citealt{rutledgeetal01a},
(5) \citealt{campanaetal04},
(6) This work
}
\label{tab:obs}
\end{deluxetable*}

\subsection{\asca}

We use the pipeline produced spectra for the \asca\ observation, obtained from HEASARC.  There is a second \asca\ observation of \cen\ taken in 1997 \citep{asai98}.  However, as noted by \citet{rutledgeetal01a} this observation is of significantly lower S/N than the first \asca\ observation and therefore adds little extra to our aim of understanding the variability of \cen.

\subsection{Chandra data reduction}

We reprocessed the data following the standard \cxo{} analysis threads, using CIAO version 4.2.  Both observations used the ACIS-S instrument.  The \verb|psextract| tool was used to extract the source and background spectra.  For both \cxo{} observations we used a circular source region with a 10 pixel radius, and for the background an annulus with inner radius of 15 pixels and outer radius of 45 pixels centered on the source.  The response files were created using the \verb|mkacisrmf| and \verb|mkarf| tools.  The source spectra were binned to a minimum of 20 counts per bin in the energy range 0.3 -- 10 keV to allow the use of $\chi^2$ statistics in spectral fitting.

\subsection{XMM-Newton data reduction}

Data were analyzed using the \xmm{} Science Analysis Software, version 9.0.0.  Calibrated event lists were created from the Observation Data Files using the latest calibration files.  The first \xmm{} observation was performed with the MOS and PN detectors operated in prime full window mode with the thin filters.  This first \xmm{} observation suffered large background flaring throughout the observation.  We therefore only extracted spectra from times with low background.  For the MOS detectors, we excluded times when the count rate from events $>10$ keV was higher than 0.5 c/s.  For the PN detector, we excluded times when the count rate from events in the range 10 -- 12 keV was higher than 1.0 c/s.  This reduced the net exposure times to 36.7, 36.5 and 23.0 ks for the MOS 1, MOS 2 and PN detectors, respectively.  We used a circular source extraction region of radius 45\arcsec{} for MOS detectors and 40\arcsec{} for the PN.  For background extraction regions we used a source-free, nearby 2\arcmin{} circular region for the MOS detectors, and two source-free, nearby circular regions of 40\arcsec{} and 60\arcsec{} for the PN.  Net count rates were 0.27, 0.29 and 1.2 c/s for the MOS 1, MOS 2 and PN, respectively.  All source spectra were binned to a minimum of 20 counts per bin.

The second \xmm{} observation was performed with the MOS and PN detectors operated in timing mode.  This second \xmm{} observation also suffered from significant background flaring, therefore we filtered the data, excluding times with high background.  For the MOS detectors we excluded times when the count rate from events $>10$ keV was higher than 0.05 c/s.  For the PN detector, we excluded times with the count rate from events in the range 10 -- 12 keV was higher than 0.5 c/s.  The net exposure times were reduced to 64.3, 63.4 and 58.9 ks for the MOS 1, MOS 2 and PN, respectively.   The source extraction regions had RAWX values of 306 -- 330, 296 -- 320 and 31 -- 45 for the MOS 1, MOS 2 and PN.  Background extraction regions were taken towards the edge of the exposed part of the detectors, away from the source, and had RAWX values 260 -- 284, 261 -- 272 and 7 -- 21 for the MOS 1, MOS 2, and PN.  Net count rates were 0.19, 0.19 and 0.86 c/s (MOS 1, MOS 2 and PN).  The background rate was higher during this observation than the first, thus significantly higher counts per bin was required to give significant detections in all bins.  We binned to a minimum of 150 counts per bin for MOS 1 and MOS 2, and 500 counts per bin for the PN.

\subsection{Suzaku data reduction}

The \suz{} observations were performed with a full window, and with normal clocking modes, with the telescope at the nominal XIS aimpoint.  The data were analyzed using HEASoft version 6.8, which includes the Suzaku v15 software.  

Data reduction for the XIS detectors, which cover the soft X-ray energy band (approximately 0.5 -- 10 keV), follows. The source spectra were extracted for each detector (XIS 0, 1 and 3) using a circular region of radius 250 px, centered on the source.  The background spectra were extracted from an annulus with inner radius 300 px and outer radius 425 px, centered on the source.  The responses were generated with the \verb|xisrmfgen| and \verb|xissimarfgen| tools.  We co-added the spectra from the two front-illuminated detectors (XIS 0 and XIS 3) to increase the signal-to-noise ratio.  The spectra we binned to a minimum of 250 counts per bin in the 0.5 -- 10 keV energy range.

We also extracted a spectrum from the hard X-ray detector PIN camera, which covers the energy range from approximately 10 -- 70 keV.  The spectrum was extracted using the \verb|hxdpinxbpi| tool which also extracts the background spectrum.  The net source exposure was 124 ks, however, the source was not detected, and the spectrum is consistent with the background.

\section{Spectral analysis}

\begin{figure}
\centering
\includegraphics[angle=270,width=8cm]{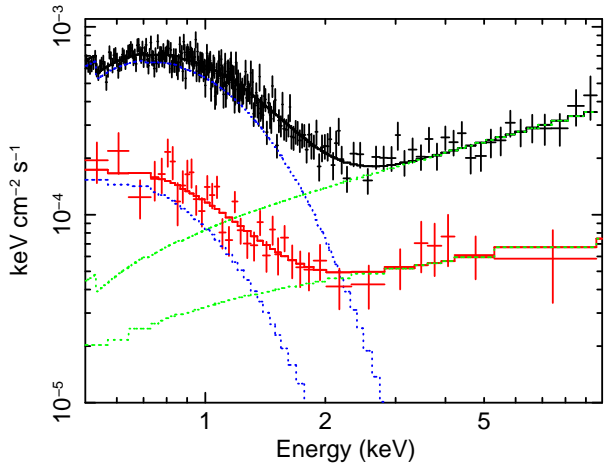}
\caption{Unfolded spectra from the brightest  (XMM1, black) and faintest (SUZ, red) quiescent observations of \cen.  The neutron star atmosphere (blue, dashed line) and power-law (green, dotted line) components are also shown.  The large variability requires that the thermal component has to vary between the two epochs.  For XMM1 we only show the PN spectrum, and for SUZ we show just the combined XIS 0 + 3 spectrum.  The spectra have been rebinned for visual clarity.}
\label{fig:unfold}
\end{figure}

The spectrum of \cen\ has clearly been shown in the past to have both a thermal and non-thermal component (see Fig.~\ref{fig:unfold}).  We therefore choose to fit the spectra with an absorbed neutron star hydrogen atmosphere plus power-law model.  We use the \verb|nsatmos| model \citep{heinke06}  which includes thermal electron conduction and self-irradiation by photons from the compact object. It also assumes a negligible magnetic field ($<10^9$ G), which is relevant here.  Spectral fitting is performed using XSpec v12 \citep{arnaud96} throughout, and all uncertainties quoted are at the $1\sigma$ level of confidence.

The aim is to investigate the variability of \cen.  Through spectral fitting we test whether the thermal component is variable, or whether changes in the absorbing column and power-law component  can explain the variability.  We therefore fit the spectra simultaneously, tying several parameters between the spectra and letting others vary freely.  We do this in several ways.  Firstly, we allow the temperature of the neutron star atmosphere to vary between observations, but hold the emitting radius fixed at 10 km. Secondly, we hold the temperature tied between the observations (i.e. the temperature is the same for all observations, but this temperature is a free parameter in the fit) but allow the emitting radius to vary.  In both cases the power-law component is left free to vary, and the photoelectric absorption column density (XSpec model \verb|phabs|) is tied between all observations.  Finally, we also test the possibility that there are changes in both the column density and the power-law while the thermal component remains unchanged.  In this case both the absorption column density, $N_{\rm H}$ and the power-law component (both the index and normalization) are free to vary between epochs, while the temperature and radius of the neutron star atmosphere component are tied between the epochs (but still a free parameter in the fit).  All spectral fits are within the energy range 0.5 -- 10 keV.

The \asca\ observations consist of four spectra (2 GIS and 2 SIS).  The parameters are tied between all the \asca\ spectra.  Similarly for the \suz\ observation we have two spectra, one from the combined XIS 0 + 3, and one from the XIS 1 detector.  Again, parameters are tied between the detectors.  For the first \xmm{} observation we have three spectra (two MOS, one PN), and tie all parameters between the detectors. When fitting the second \xmm{} observation on its own, we found a offset between the MOS and PN detectors (which were both operated in timing mode).  Fitting the absorbed neutron star atmosphere plus power-law model to all three spectra with all parameters tied between the detectors led to a poor fit ($\chi_\nu^2 = 2.81$).  However, if a constant offset was used an acceptable fit was achieved ($\chi_\nu^2=1.02$).  The constant was fixed to 1.0 for the PN, and the best-fitting value for the MOS spectra was $1.26\pm0.01$ (the constant was tied between the MOS 1 and MOS 2).  Comparing fitting the PN alone and fitting all 3 spectra with a constant, we found only very minor differences in the fit parameters.  Thus, in the remaining analysis, we fit all 3 spectra with a constant offset between the PN and MOS to achieve the best constraints on parameters.

Spectral fit results from allowing the temperature to vary are given in Table~\ref{tab:Tfree_Rfix}, spectral fit results from allowing the radius to vary are given in Table~\ref{tab:Tfix_Rfree}, and spectral fits from allowing both the column density and power-law to vary are given in Table~\ref{tab:NHvariable}.

\tabletypesize{\footnotesize}
\begin{deluxetable*}{lcccccc}
\tablecolumns{7}
\tablewidth{0pc}
\tablecaption{Spectral fits for \cen, with temperature variable and radius fixed}
\tablehead{ & ASCA & CXO1 & XMM1 & XMM2 & CXO2 & SUZ}
\startdata
$N_{\rm H}$ ($10^{20}$ cm$^{-2}$) & \multicolumn{6}{c}{$4.9\pm0.2$} \\
$kT_{\rm eff}^{\infty}$ (eV)      & $63.3\pm0.7$  & $59.2\pm0.4$  & $66.2\pm0.2$  & $62.0\pm0.2$  & $51.2\pm0.6$  & $48.2\pm0.6$  \\
Power-law index, $\Gamma$         & $1.24\pm0.17$ & $0.97\pm0.25$ & $1.41\pm0.05$ & $1.26\pm0.08$ & $0.78\pm0.43$ & $1.69\pm0.17$ \\
Power-law norm. ($10^{-5}$)       & $7.9\pm1.6$   & $3.1\pm0.9$   & $9.4\pm0.6$   & $5.4\pm0.5$   & $1.4\pm0.7$   & $3.7\pm0.6$ \\
Unabs. 0.5 -- 10 keV flux ($10^{-12}$ erg s$^{-1}$ cm$^{-2}$)         & $1.86\pm0.17$ & $1.20\pm0.13$ & $2.09\pm0.06$ & $1.48\pm0.03$ & $0.63\pm0.10$ & $0.47\pm0.03$ \\
Unabs. 0.5 -- 10 keV thermal flux ($10^{-12}$ erg s$^{-1}$ cm$^{-2}$) & $1.00\pm0.09$ & $0.71\pm0.07$ & $1.26\pm0.04$ & $0.90\pm0.02$ & $0.33\pm0.06$ & $0.24\pm0.02$ \\
$\chi_\nu^2$ (dof) & \multicolumn{6}{c}{0.99 (1166)}
\enddata
\tablecomments{The neutron star atmosphere model `nsatmos' is used. For all observations, the neutron star radius was fixed at 10 km, and the mass at 1.4 M$_\odot$.  The distance was fixed at 1.2 kpc. $N_{\rm H}$ was tied between all observations. $kT_{\rm eff}^{\infty}$ is the effective temperature for an observer at infinity, for $R = 10$ km and $M = 1.4 $M$_\odot$. The power-law normalization is defined as photons keV$^{-1}$ cm$^{-2}$ s$^{-1}$ at 1 keV.}
\label{tab:Tfree_Rfix}
\end{deluxetable*}

\begin{deluxetable*}{lcccccc}
\tablecolumns{7}
\tablewidth{0pc}
\tablecaption{Spectral fits for \cen, with temperature tied and radius variable}
\tablehead{ & ASCA & CXO1 & XMM1 & XMM2 & CXO2 & SUZ}
\startdata
$N_{\rm H}$ ($10^{20}$ cm$^{-2}$) & \multicolumn{6}{c}{$5.6\pm0.6$} \\
$kT_{\rm eff}$ (eV)      & \multicolumn{6}{c}{$79.3\pm2.5$} \\
Radius (km)                       &  $10.8\pm0.6$ & $9.7\pm0.6$   & $11.7\pm0.8$  & $10.5\pm0.5$ & $7.8\pm0.4$ & $7.2\pm0.3$ \\
Power-law index, $\Gamma$         & $1.29\pm0.11$ & $0.85\pm0.14$ & $1.51\pm0.04$ & $1.22\pm0.05$ & $0.31\pm0.16$ & $1.53\pm0.10$ \\
Power-law norm. ($10^{-5}$)       & $8.5\pm1.7$   & $2.6\pm0.8$   & $11.1\pm0.8$  & $5.1\pm0.5$ & $0.8\pm0.5$ & $3.0\pm0.6$ \\
Unabs. 0.5 -- 10 keV flux ($10^{-12}$ erg s$^{-1}$ cm$^{-2}$)         & $1.88\pm0.15$ & $1.24\pm0.12$ & $2.13\pm0.05$ & $1.51\pm0.05$ & $0.72\pm0.10$ & $0.48\pm0.04$ \\
Unabs. 0.5 -- 10 keV thermal flux ($10^{-12}$ erg s$^{-1}$ cm$^{-2}$) & $1.01\pm0.08$ & $0.74\pm0.07$ & $1.27\pm0.03$ & $0.94\pm0.03$ & $0.35\pm0.05$ & $0.25\pm0.02$ \\
$\chi_\nu^2$ (dof) & \multicolumn{6}{c}{0.99 (1165)}
\enddata
\tablecomments{The neutron star atmosphere model `nsatmos' is used. For all observations, the mass was fixed at 1.4 M$_\odot$.  The distance was fixed at 1.2 kpc. Both $N_{\rm H}$ and $kT_{\rm eff}$ were tied between all observations. Note that here we give the unredshifted $kT_{\rm eff}$ as the emitting radius (and thus the redshift) is allowed to be different at each epoch. The power-law normalization is defined as photons keV$^{-1}$ cm$^{-2}$ s$^{-1}$ at 1 keV.}
\label{tab:Tfix_Rfree}
\end{deluxetable*}
 
\begin{deluxetable*}{lcccccc}
\tablecolumns{7}
\tablewidth{0pc}
\tablecaption{Spectral fits for \cen, with variable $N_{\rm H}$ and power-law}
\tablehead{ & ASCA & CXO1 & XMM1 & XMM2 & CXO2 & SUZ}
\startdata
$N_{\rm H}$ ($10^{20}$ cm$^{-2}$) &  $12.0\pm1.4$ & $14.6\pm0.8$ & $10.0\pm0.6$ & $11.6\pm0.7$ & $31.5\pm1.5$ & $44.9\pm1.7$ \\
$kT_{\rm eff}^{\infty}$ (eV)      & \multicolumn{6}{c}{$48.7^{+0.5}_{-0.9}$} \\
Radius (km)                       &  \multicolumn{6}{c}{$25.7\pm0.5$} \\
Power-law index, $\Gamma$         & $2.0\pm0.1$ & $1.3\pm0.1$ & $2.4\pm0.1$ & $1.9\pm0.1$ & $-1.0^{+0.4}_{-0.7}$ & $0.70\pm0.14$ \\
Power-law norm. ($10^{-5}$)       & $20.1\pm2.2$   & $5.1\pm1.0$   & $32.1\pm1.2$  & $13.1\pm0.9$ & $0.16^{+0.15}_{-0.09}$ & $0.87\pm0.23$ \\
Unabs. 0.5 -- 10 keV flux ($10^{-12}$ erg s$^{-1}$ cm$^{-2}$)  & $2.10\pm0.09$ & $1.61\pm0.11$ & $2.31\pm0.04$ & $1.78\pm0.03$ & $1.99^{+0.07}_{-1.51}$ & $1.32^{+0.14}_{-0.54}$ \\
$\chi_\nu^2$ (dof) & \multicolumn{6}{c}{1.47 (1165)}
\enddata
\tablecomments{The neutron star atmosphere model `nsatmos' is used. For all observations, the mass was fixed at 1.4 M$_\odot$.  The distance was fixed at 1.2 kpc. Both the column density, $N_{\rm H}$, and the power-law (index and normalization) were allowed to vary between epochs, while the neutron star effective temperature and radius were tied between all observations. $kT_{\rm eff}^{\infty}$ is the effective temperature for an observer at infinity, for $M = 1.4 $M$_\odot$ and the best-fitting radius. The power-law normalization is defined as photons keV$^{-1}$ cm$^{-2}$ s$^{-1}$ at 1 keV.}
\label{tab:NHvariable}
\end{deluxetable*}

\section{Lightcurves}

\begin{figure*}
\centering
\includegraphics[angle=270,width=13cm]{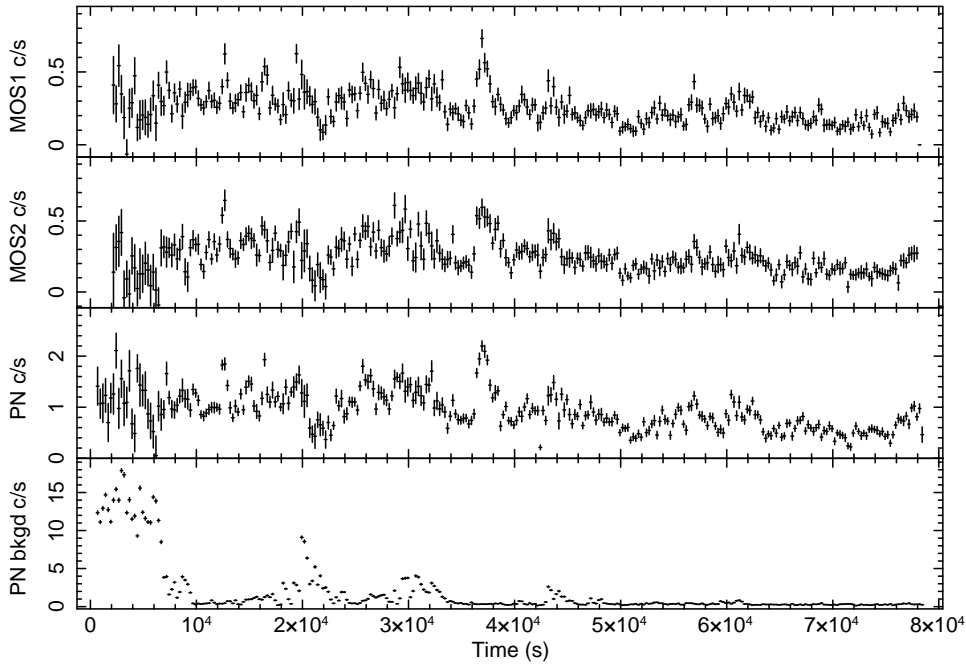}
\caption{Lightcurves from the second \xmm{} observation (XMM2).  The top three panels show the MOS 1, MOS 2 and PN net count rate, with 250 s binning in the 0.3 -- 10 keV energy range.  The bottom panel shows the PN background count rate (note that periods of high background excluded from the spectral analysis have not been filtered out here).  While some variability in the net lightcurves may be associated with higher levels of background (for instance at around 20 ksec), there is a clear flare from the source at around 37 ksec seen in all three detectors during a period of low background.}
\label{fig:xmmlc}
\end{figure*}

\begin{figure}
\centering
\includegraphics[angle=270,width=8cm]{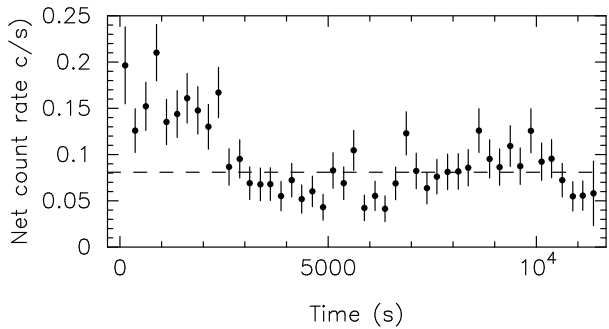}
\caption{0.3 -- 10 keV lightcurve from the second \cxo{} observation (CXO2), with 250 s binning. The dashed line is the weighted average of the lightcurve.  The early section of the lightcurve is significantly higher than the average.}
\label{fig:cxo2lc}
\end{figure}

In addition to studying the long-term variability of \cen{} from epoch to epoch, we also take a look at the variability during the observations newly analyzed here (XMM2, CXO2, SUZ).  We extracted background-subtracted lightcurves using the same source and background regions as for the spectral analysis.  Given the low count rate, the lightcurve from the \suz{} observation is noisy, and does not show any clear variability.  However, the lightcurves from both XMM2 and CXO2 show some short-term variability.  The lightcurves from XMM2 are shown in Fig.~\ref{fig:xmmlc}, and the lightcurve from CXO2 is shown in Fig.~\ref{fig:cxo2lc}.

The XMM2 lightcurves were extracted from the entire data, including the time periods that were excluded from the spectral analysis due to high background flaring.  We show the background-subtracted 0.3 -- 10 keV lightcurves with 250-s binning from all three EPIC detectors, as well as showing the background lightcurve from the PN.  Some of the variability seen in the lightcurve may be associated with imperfect background subtraction during periods of high background flaring.  A clear example of this is at around 20 ksec.  However, there is a significant flare in the source lightcurves at around 37 ksec that is during a period of low background, thus is clearly associated with source variability.

\begin{figure}
\centering
\includegraphics[angle=270,width=7cm]{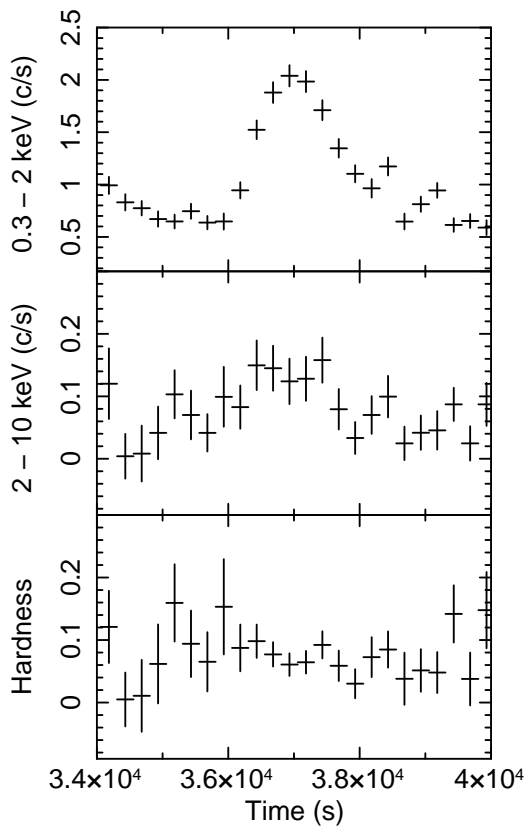}
\caption{The XMM2, PN lightcurves (with 250 s binning) around the time of a significant flare.  Here we show the 0.3 -- 2 keV (top) and 2 -- 10 keV (middle) lightcurves, and the associated hardness ratio (2 -- 10 keV count rate / 0.3 -- 2 keV count rate).  There is no obvious evolution in the hardness ratio during the flare.}
\label{fig:flare}
\end{figure}

In order to investigate the nature of this flare, we extracted lightcurves in the 0.3 -- 2.0 keV range and the 2.0 -- 10 keV range from the PN data, and looked at the hardness ratio (2.0 -- 10 keV count rate/ 0.3 -- 2.0 keV count rate) for any significant changes (see Fig.~\ref{fig:flare}).  The flare is prominent in the soft lightcurve, though there is no significant evolution in the hardness ratio during the flare.  It is therefore not possible to conclude which component caused the flare.

We perform a simple test for variability in the CXO2 observation by fitting a constant to the lightcurve.  A constant provides a poor fit ($\chi_\nu^2 = 3.0$), with the early part of the lightcurve significantly above the weighted average (see~Fig.~\ref{fig:cxo2lc}), indicating variability on a timescale of a few hundred seconds.

\section{Results}

The spectral analysis clearly shows a large amplitude of variability between epochs, with the flux varying by a factor of 4.4 between the brightest and faintest observations (XMM1 and SUZ, respectively).  We show the unfolded  spectra from these two observations in Figure~\ref{fig:unfold}, and the spectral fits shown are for varying temperature and fixed radius.  What is clear from the spectral fits is that the thermal component has to have varied between these two epochs.  Given that the power-law accounts for approximately 40\% of the 0.5 -- 10 keV flux in XMM1, a factor of 4.4 change in flux cannot be achieved through a variable power-law alone.  The spectral fits allowing both the power-law and column density to vary also do not fit the data well, leading to a significantly worse fit ($\chi_\nu^2 = 1.47$, $\nu = 1165$) than when allowing the thermal component to be variable.

A spectral fit with the temperature tied between all observations, and the radius fixed at 10 km is statistically not acceptable ($\chi_\nu^2 = 1.91$, $\nu = 1171$).  The brightest observations all have very soft power-law indices ($\sim3$), and poor fits above 3 keV, as the power-law component tries to fit the majority of the thermal component.  We also tried fitting with the radius as a free parameter, though having its value be the same for all observations, however this does not significantly improve the fit ($\chi_\nu^2 = 1.90$, $\nu = 1170$).  Thus, the spectral fits demonstrate that the thermal component must vary between epochs, and the variability cannot be attributed to changes in the power-law index, normalization and/or the absorption.

\begin{figure}
\centering
\includegraphics[angle=270,width=7cm]{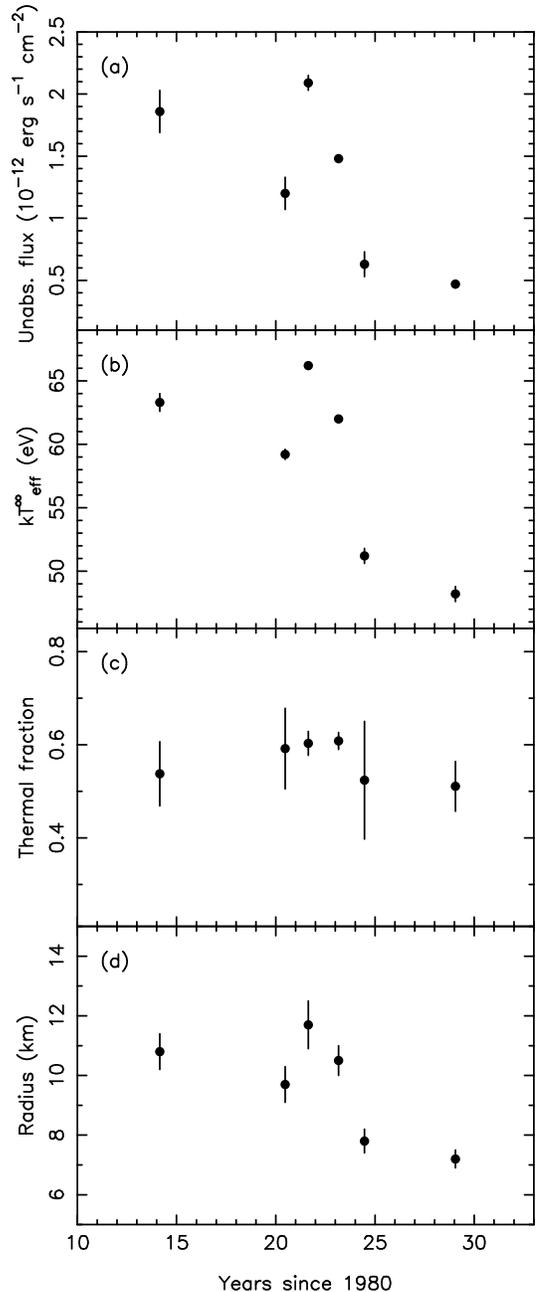}
\caption{(a) Long-term lightcurve of \cen.  The data shows the unabsorbed 0.5 -- 10 keV flux from fitting with the temperatures tied and a radius = 10 km. (b) Effective temperature (for an observer at infinity) vs. time.  In the spectral fits the neutron star radius was fixed at 10 km.  (c) Thermal fraction vs. time, where the thermal fraction is defined as the ratio of the unabsorbed 0.5 -- 10 keV neutron star atmosphere flux to the total unabsorbed 0.5 -- 10 keV flux. This is using the spectra fits with the temperature tied and a radius = 10 km. (d) Neutron star radius vs. time from spectral fits assuming that the neutron star atmosphere temperature is the same at all epochs and that the radius varies.}
\label{fig:longterm}
\end{figure}

Figure \ref{fig:longterm} shows the long-term variability of \cen{}.  The top three panels in this figure show the variability of the (a) unabsorbed 0.5 -- 10 keV flux, (b) effective temperature and (c) fraction of flux in the thermal component from the spectral fits where the temperature is variable between epochs, and the radius fixed.  Both the changes in flux and temperature follow the same overall pattern.  In fact, it can be seen that the thermal fraction remains consistent between epochs, implying that the flux from the thermal and power-law components is varying by approximately the same amount.  The bottom panel, (d), shows how the radius changes in the case where the temperature is tied between each epoch in the spectral fits.  It demonstrates that in this model, significant changes in the radius are required to account for the spectral changes.  Therefore, to achieve the observed flux variability, either the temperature of the thermal component, or the emitting radius must vary between epochs.  We cannot rule out that both the temperature {\it and} the radius change.

For the sake of completeness, we also note several previous quiescent observations not studied here. {\it Einstein} and {\it EXOSAT} observations were performed in 1980 and 1986 respectively \citep{vanparadijsetal87}, though their flux measurements are not precise enough to be constraining here \citep[see table 2 from][which concisely summarizes these observations]{rutledgeetal01a}.  {\it ROSAT}/HRI observed \cen{} in 1995 \citep{campana97}, however, given that this was not a spectral instrument no tight flux constraint is obtained.  As noted earlier, there was a second {\it ASCA} observation of \cen{} in quiescence \citep{asai98}, which was unfortunately of low quality.  This observation was performed on February 4--5, 1997.  The analysis by \citet{asai98} led to a rather poorly constrained unabsorbed 0.5 -- 10 keV flux of  $1.75^{+1.75}_{-1.17}\times10^{-12}$ erg s$^{-1}$ cm$^{-2}$, consistent with all but our faintest observation.  Finally, a {\it BeppoSAX} observation was performed on February 9, 1999 \citep{campana00}, and this was found to be consistent with the first {\it Chandra} observation \citep{rutledgeetal01a}.

\section{Discussion}

The two known outbursts from the neutron star LMXB \cen{} were in 1969 and 1979, and the source has been in quiescence ever since (it has not been detected in outburst by any mission or all-sky monitor).  Our spectral analysis of six observations during quiescence covers a period of over 15 years, and shows variability of a factor 4.4 between the brightest and faintest states.  In all observations both a thermal and power-law component is present in the spectrum.  The amplitude of the observed variability requires that the power-law alone cannot account for the variability, as the power-law component only contributes about 40\% of the flux during the brightest state.  Our spectral fitting demonstrates that the thermal component must vary with either the temperature and/or the radius of the thermal component changing between epochs.  Allowing both the power-law component and the column density to vary (while the thermal component remains constant) does not fit the data well.  

In addition to the long-term epoch to epoch variability, we have observed short-term variability on timescales of the order of a few hundred seconds in the lightcurves from two of the newly analyzed datasets (XMM2 and CXO2).  Such short-term variability has previously been observed in \cen{} \citep{campanaetal04}, though the nature (whether thermal or power-law) was inconclusive.  Given the source was fainter during these two new observations than during the study by \citet{campanaetal04}, we again are unable to determine the cause of this short-term variability.

There are several methods by which the temperature of the neutron star surface may change on timescales of years, which we now discuss in turn.   Firstly, once a source returns to quiescence, the crust may cool \citep[e.g.][]{rutledge_ks1731_02}.  In the case where the outburst is particularly long (i.e., lasting several years rather than several weeks to months), the crust can be heated significantly out of thermal equilibrium with the rest of the star.  Therefore, once accretion reduces to quiescent levels, the crust will thermally relax.  Such crustal cooling has been observed in four sources so far.  It was first observed in KS~1731$-$260 and MXB~1659$-$29 which both returned to quiescence in 2001 \citep{wijnands01,wijnands03}.   Since then, monitoring with {\it Chandra} and {\it XMM-Newton} has shown both sources to cool rapidly \citep{wijnandsetal02,wijnandsetal04,cackett06}.  The cooling curve of MXB~1659$-$29 covers 6.6 years, and is well described by an exponential decay to a constant level, with a $e$-folding timescale of $465\pm25$ days \citep{cackett06,cackett08}.  While the cooling curve of KS~1731$-$260 can also be described by exponential cooling to a constant, with a similar $e$-folding time \citep{cackett06}, the most recent observation suggests that cooling is continuing following a power-law decay (Cackett et al., in preparation). 

In the last couple of years, two additional sources with long outbursts have also gone into quiescence \citep{degenaar09,fridriksson10}.  The neutron star transient EXO 0748$-$676 was in outburst before returning to quiescence.  {\it Chandra} and {\it Swift} observations covering the first 5 months after the source transitioned to quiescence shows an initially slow decrease in temperature \citep{degenaar09}, with further observations ongoing.  Finally, once the transient XTE~J1701$-$462 returned to quiescence it displayed rapid cooling \citep{fridriksson10}.  However, the cooling curve observed is complicated by a temporary increase in temperature about 220 days into quiescence.   After this increase the cooling continued on the same track as before the increase -- an exponential decay to a constant, with an $e$-folding time of about 120 days.   \citet{fridriksson10} suggest that the apparent increase in temperature could be caused by an increased level of accretion.

The long-term quiescent lightcurve for \cen{} (Fig.~\ref{fig:longterm}) may show an overall decrease suggestive of cooling,  but with several increased points, as observed in XTE~J1701$-$462.   Clearly, though, the sampling rate is extremely infrequent and there are only a small number of points.   However, the timescale of any apparent decrease is significantly longer than the $e$-folding timescales for the crustal cooling sources discussed above. The variability we observe from \cen{} is from 15-30 yr after the 1979 outburst, and thus the crust should have cooled back into thermal equilibrium with the core at this point if it has a similar structure to the other sources.  It therefore seems unlikely that crustal cooling contributes significantly to the variability observed.  Note that the neutron star core will not cool appreciably over the timescales observed here \citep[e.g.,][]{yakovlev04} and thus will not contribute to any variability.

Continued low-level accretion onto the neutron star surface can also change the thermal properties of the star.  The thermal quiescent flux is sensitive to both the amount of H/He remaining on the surface after an outburst and on the composition of ashes from previous H/$^4$He burning \citep*{BBC02}.  Continued low-level accretion will change the surface composition by adding an insulating layer of ashes from H/$^4$He burning.  This therefore changes the surface effective temperature, and can occur on timescales of $> 10$ yr at the luminosity of \cen.  However, this effect should lead to an increase in temperature over time.  Specific calculations for \cen{} by \citet{BBC02} suggest only a 20\% increase in brightness over 30 years due to this effect.  Here, we see variations greater than 20\% and variability that is not just a simple increase, suggesting that an increase in the depth of the H/$^4$He layer cannot explain the observed variability in \cen.

Another process by which the thermal properties of a quiescent neutron star may vary is diffusive nuclear burning \citep{chang03}.  In this process, hydrogen diffuses to deeper layers in the crust where it can fuse.  This would lead to a drop in flux as hydrogen is consumed and the hydrogen abundance in the photosphere decreases.  This process is only expected to lead to a changes on timescales $> 10$ yr.  For \cen{}, \citet{chang03}  specifically predict that 20 yr after the outburst the flux will vary by only 3\% over a 10 yr timescale, and that roughly 100 yr after the outburst the flux will drop by 12\% over a 20 yr period.  Consequently, it is hard to explain the observed variability by this process, given that we see variability of much greater amplitude than this, as well as more than just a simple flux decline.

The above explanations are also disfavored because the thermal fraction is constant while the luminosity varies.  However, one other possibility is that variable residual accretion is responsible for the observed behavior.  If low-level accretion onto the neutron star surface is causing the power-law component in the spectrum, then an increase in this accretion rate would increase the power-law, which could in turn lead to an increase in the observed neutron star temperature.  In fact, residual accretion is known to be able to produce a thermal spectral shape \citep{zampierietal95}.  One striking finding is that  while the total luminosity varies by a factor of over 4 the ratio of thermal to power-law flux remains approximately constant (Fig.~\ref{fig:longterm}, panel c).  This appears to  indicate that the thermal and power-law components are linked together, thus any mode for residual accretion would need to produce both components and in a way that their ratio remains constant. This may be able to provide constraints on models of accretion flows at low mass-transfer rates. 

In summary, the crustal cooling, H/He ashes and diffusive nuclear burning processes all seem unlikely to drive the thermal variability observed, and the most likely scenario appears to be one where variable low-level accretion causes both the thermal and non-thermal variability.  More frequent monitoring of \cen{} will lead to a better understanding of both the amplitude and timescales of the variability.  Finally, this thermal variability may pose a problem for measuring neutron star radii from modeling quiescent emission given that it is not clear whether the emitting radius, temperature or both is causing the variability.  If the variability is due to low-level accretion, then avoiding sources with a significant power-law component would be beneficial.

\acknowledgements
EMC gratefully acknowledges support provided by NASA through the Chandra Fellowship Program, grant number PF8-90052.  

\bibliographystyle{apj}
\bibliography{apj-jour,qNS}

\end{document}